\documentclass[singlecolumn,showpacs,amsmath,amssymb,amsfonts,nofootinbib,plb]{revtex4}
\usepackage{graphicx}
\usepackage{dcolumn}
\usepackage{bm}
\usepackage{ulem}
\usepackage{overpic}
\usepackage{amssymb}

\newcommand{\nn}{\nonumber}

\def\beq{\begin{equation}}
\def\eeq{\end{equation}}

\usepackage{soul}
\usepackage{cancel}

\usepackage[usenames]{color}

\usepackage{graphicx}
\usepackage{dcolumn}
\usepackage{bm}
\usepackage{ulem}
\usepackage{overpic}
\usepackage{amssymb}
\usepackage{amsmath}
\usepackage{datetime}
\usepackage{xcolor}
\usepackage{arydshln}

\usepackage{color}
\definecolor{myblue}{rgb}{.8, .8, 1}
\definecolor{apricot}{rgb}{0.98, 0.81, 0.69}

\usepackage{siunitx,amsmath}
\usepackage{amsmath}
\usepackage{empheq}

\usepackage{hyperref}
\hypersetup{colorlinks=true,
linkcolor=blue,
filecolor=magenta,
urlcolor=black,}

\newlength\mytemplen
\newsavebox\mytempbox
\makeatletter
\newcommand\mybluebox{%
    \@ifnextchar[
       {\@mybluebox}%
       {\@mybluebox[0pt]}}

\def\@mybluebox[#1]{%
    \@ifnextchar[
       {\@@mybluebox[#1]}%
       {\@@mybluebox[#1][0pt]}}

\def\@@mybluebox[#1][#2]#3{
    \sbox\mytempbox{#3}%
    \mytemplen\ht\mytempbox
    \advance\mytemplen #1\relax
    \ht\mytempbox\mytemplen
    \mytemplen\dp\mytempbox
    \advance\mytemplen #2\relax
    \dp\mytempbox\mytemplen
    \colorbox{myblue}{\hspace{1em}\usebox{\mytempbox}\hspace{1em}}}

\makeatother

\def\beq{\begin{equation}}
\def\eeq{\end{equation}} 
\usepackage{soul}
\usepackage{cancel}
\usepackage{breqn}
\usepackage{tikz}

\makeatletter
\let\cat@comma@active\@empty
\makeatother

\begin{document}
\input{epsf}

\input{epsf}

\title{Saturation Correction to Projectile - some  ${\cal{O}}(g^5)$ results within LCPT}

\author{Nahid Vasim}
\email{nvasim@myamu.ac.in}

\affiliation{Department of Physics, Aligarh Muslim University, Aligarh - $202002$, India.}

 \begin{abstract}
We revisited the first saturation correction to projectile in nucleus-nucleus collisions \cite{Chirilli:2015tea} within diagrammatic light cone perturbation theory (LCPT). To get an analytic expression for the saturation correction in projectile authors in \cite{Chirilli:2015tea} calculated complete ${\cal{O}}(g^3)$ amplitudes, where the main results are given in transverse coordinate space. However to find the complete first saturation correction in the projectile, one needs to calculate both ${\cal{O}}(g^3)$ and ${\cal{O}}(g^5)$ gluon production amplitudes. We here present a detailed calculation of a sample graph at ${\cal{O}}(g^5)$ in diagrammatic light cone perturbation theory.   
 \end{abstract}

\pacs{12.38.-t}

\date{\today}
\maketitle

\section{Introduction}

Heavy-ion collision and subsequent avalanche in the gluon production is far complex dynamical system than proton-ion or electron-ion collision. In the formal study of gluon production in the heavy-ion collision one need to resums the power of $\alpha_s^2 A_1^{1/3}$
and $\alpha_s^2 A_2^{1/3}$ \cite{Kovchegov:1997pc} where $A_1$ and $A_2$ are the atomic numbers of the two nuclei and $\alpha_s$ is strong coupling. In the saturation framework \cite{Gribov:1983ivg,Iancu:2003xm,Jalilian-Marian:2005ccm,Weigert:2005us,Gelis:2010nm,Albacete:2014fwa,Balitsky:2001gj,Kovchegov:2012mbw} the `saturation scale' also scales with the parameter $\alpha_s^2 A^{1/3}$. This leads to ansatz for the single gluon production cross-section in heavy-ion collision as, 
\begin{eqnarray}
\frac{d\sigma}{d^2k d^2B d^2b} = \frac{1}{\alpha_s}
 ~ f\left(\frac{Q_{s1}^2(B_\perp - b_\perp)}{k_\perp^2}, \frac{Q_{s2}^2(b_\perp)}{k_\perp^2}\right)~.
\end{eqnarray}
The overall $1/\alpha_s$ implies that it is on the verge of onset of non-perturbative effects and also that formulations are within the classical approximations. 
The function $f$ is essentially a Cauchy product of two infinite series in the power of  
$\alpha_s^2 A_1^{1/3}$
and $\alpha_s^2 A_2^{1/3}$  or equivalently in $Q_{s1}/k_\perp^2$ and 
$Q_{s2}/k_\perp^2$, and therefore can be written as, 
 \begin{eqnarray}
 f = \sum_{n,m = 1}^{\infty} C_{n,m} \left(\frac{Q_{s1}^2}{k_\perp^2}\right)^m
 \left(\frac{Q_{s2}^2}{k_\perp^2}\right)^n ~.
 \end{eqnarray}
In this work we have calculated a sample diagram at ${\cal O}(g^5)$ that contribute to the determination of the coefficients $c_{2,n}$ for all $n$. This is the case where one has to expand the gluon production cross section to the second order in $Q_{s1}^2$ but in all order in $Q_{s2}^2$. This demonstrate a senario where two nucleons from the projectile would interact with the all nucleons of the target.  

\vspace{0.3cm}
First saturation correction to the single inclusive gluon production was first studied by Balitsky in the context of interaction of two shock waves in QCD \cite{Balitsky:2004rr}. This was later also independently studied by Chirilli, Kovchegov and Wertephy using the diagrammatic approach based on light cone perturbation theory\cite{Chirilli:2015tea}. Later complete first saturation correction to the single inclusive semi hard gluon production was obtained by Li and Skokov from the analytical solution of classical Yang-Mills equation in the field of dilute objects by applying LSZ reduction \cite{Li:2021zmf}. Both the ${\cal O} (g^3)$ and ${\cal O} (g^5)$ amplitudes contribute to the first saturation correction to the single gluon production. Both the order  ${\cal O} (g^3)$ amplitude squared and the crossing terms between ${\cal O} (g)$ and ${\cal O} (g^5)$ are proportional to $g^6 N^4$ where $N$ is the number of color sources in the projectile. Specifically in this work we have calculated scattering amplitude of a sample diagram at ${\cal O} (g^5)$. Situation where one nucleon interact both in amplitude and the complex conjugate amplitude while other interact only in the complex conjugate amplitude would lead to contribute to the cross section $\sim$ $g^5 g$  $\sim$ $\alpha_s^3$. 
In this study we use the approach of Ref. \cite{Chirilli:2015tea} based on diagrammatic light cone perturbation theory. In Ref. \cite{Chirilli:2015tea} amplitude of all diagrams at ${\cal O} (g^3)$ was calculated. In this work we report amplitude of one sample diagram at ${\cal O} (g^5)$.  Considering the depth of involvement in calculating the diagram at ${\cal O} (g^5)$,  we hope to able to prepare a complete list of results for ${\cal O} (g^5)$  in LCPT approach elsewhere in future.

 \section{Sample diagram for gluon production amplitude at order-$g^5$ }

\begin{figure}[ht]
\begin{center}
\includegraphics[width=0.5 \textwidth]{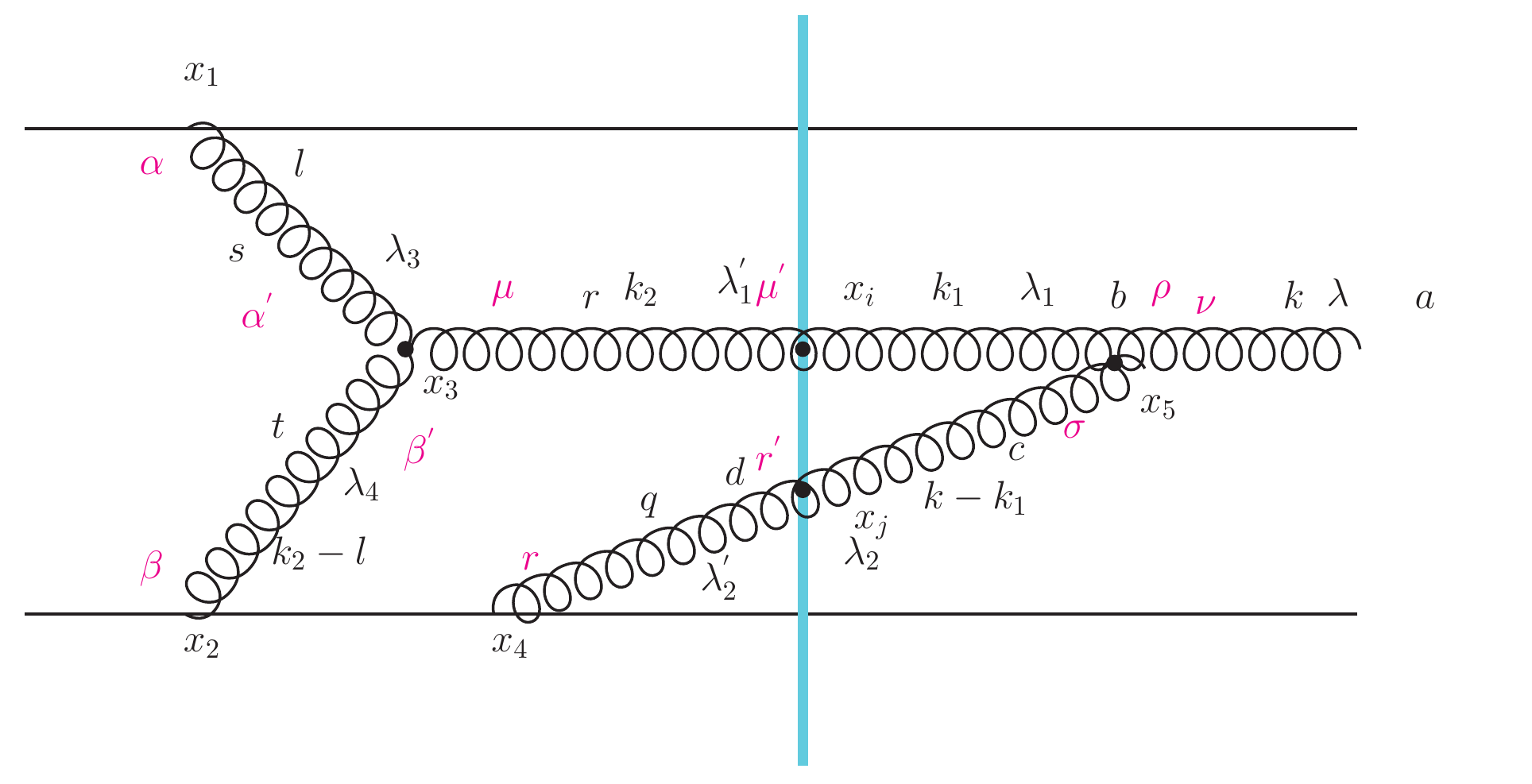} 
\caption{A detailed diagram of amplitude at order-$g^5$.}
\label{myfig}
\end{center}
\end{figure}

In the following we calculate an amplitude for gluon production at order-$g^5$.
We show a detailed calculation of a graph where all the gluons are produced before there is any interaction with the target shock wave, see Fig.\ref{myfig}. All the momentum, coordinate, polarization, Lorentz-index and color labels are distinctly shown in the figure.
The diagram contain two triple-gluon vertex and all gluon propagators are retarded, with the time flowing in the direction of the produced gluon. 
Two quarks coming from two different nucleons in the projectile is given by two Wilson lines shown as two solid horizontal lines in the diagram.
 %
 %
We need to calculate the amplitude of the diagram in the coordinate space. It is done so because in the MV model calculating the correlators of Wilson lines in transverse coordinate space is relatively easier. The two quarks from two different nucleons in the projectile moving in $x^+$-direction, have different $x^-$ positions. The coordinates are defined below:

 \begin{eqnarray}
  &&{x_1}\equiv(x_1^+,x_1^-,{x_1}_\perp) ~~~ [{x_1}^+ \leqslant 0],\nn\\
  &&{x_2}\equiv(x_2^+,x_2^-,{x_2}_\perp) ~~~ [{x_2}^+ \leqslant 0],\nn\\
  &&{x_3}\equiv(x_3^+,x_3^-,{x_3}_\perp) ~~~ [{x_3}^+ \leqslant 0],\nn\\
  &&{x_4}\equiv(x_4^+,x_4^-,{x_4}_\perp) ~~~ [{x_4}^+ \leqslant 0],\nn\\
 && {x_5}\equiv(x_5^+,x_5^-,{x_5}_\perp) ~~~ [{x_5}^+ \leqslant 0].
 \end{eqnarray}
All gluon propagators are retarted Green function, such that time flows towards the measured gluon in the diagram. The time-ordering of these propagators is such that the gluons are emitted first,
and then they participate in an interaction.
Thus we see gluons are emitted from two quark lines at $x_1$ and $x_2$. They meet at first triple gluon  vertex, $x_3$.  An other gluon emitted from quark line  at $x_4$ and the gluon emitted at $x_3$, first interact with the shock wave then meet at other triple gluon vertex leading to production of measured gluon. 
 The numerator of gluon propagators are written in terms of polarization sum: \\
\begin{eqnarray}
 D^{\alpha\alpha^'}(l)&=&\sum_{\lambda_3}\epsilon_{\lambda_3}^{*\alpha}(l)\epsilon_{\lambda_3}^{\alpha^'}(l),
 \end{eqnarray}
In the $A^+=0$ gauge the polarization vector $\epsilon_{\lambda}^{\alpha}(k)$ is, 
\begin{eqnarray}
 \epsilon^\mu (l) = \left( 0, \frac{{\vec \epsilon}_\perp^{\,
        \lambda} \cdot {\vec l}_\perp}{l^+}, {\vec \epsilon}_\perp^{\,
      \lambda} \right),
\end{eqnarray}
and the two 3-gluon vertices are \cite{Kovchegov:2012mbw}:
\begin{eqnarray}
 &&-gf^{srt}\left[(2l-k_2)_{\mu}g_{\alpha^'\beta^'}-(k_2+l)_{\beta^'}g_{\alpha^'\mu}+(2k_2-l)_{\alpha^'}g_{\mu\beta^'}\right]\nn\\
 &&-gf^{bac}\left[(2k_1-k)_{\nu}g_{\rho\sigma}-(k+k_1)_{\sigma}g_{\rho\nu}+(2k-k_1)_{\rho}g_{\nu\sigma}\right]
\end{eqnarray}
We separately calculated entire numerator for the amplitude in Appendix \ref{Num}.\\
\vspace{0.5cm}

In the calculation the quark lines are treated as light cone Wilson lines.
According to standard rules of the eikonal approximation \cite{Balitsky:1995ub} the gluon interaction with the shock wave brings in a factor of $(2k_1^+)U^{br}_{x_{5\perp}}$ for a gluon line with light-cone momentum $k_1^+$ and transverse
coordinate $x_{5\perp}$, and similarly interaction of a quark with a shock wave yields $(V_{x_{1\perp}}t^s)_1$ for a quark line with transverse coordinate~$x_{1\perp}$. The amplitude for the sample diagram is written as,

\begin{eqnarray}
A_1&=&(ig)^3\int_{-\infty}^{0}dx_1^+~e^{\epsilon x_1^+}~\int_{-\infty}^{0}dx_2^+~e^{\epsilon x_2^+}~\int_{-\infty}^{0}
dx_3^+~e^{\epsilon x_3^+}\int_{-\infty}^{0}dx_4^+~e^{\epsilon x_4^+}~\int_{0}^{\infty}dx_5^+~e^{\epsilon x_5^+}~e^{ik^-x_5^+}\nn\\
&& (2k_1^+)U^{br}_{x_{5\perp}}~\left(2(k-k_1)^+\right)U^{cd}_{x_{4\perp}} (V_{x_{1\perp}}t^s)_1(V_{x_{2\perp}}t^dt^t)_2\nn \\
&& \int\frac{dl^+}{2\pi}e^{-il^+(x_3^--x_1^-)}   \int\frac{d(k_2-l)^+}{2\pi}e^{-i(k_2-l)^+(x_3-x_2)^-}
 \int\frac{dk_2^+}{2\pi}e^{-ik_2^+(x_i^--x_3^-)}  \int\frac{dq^+}{2\pi}e^{-iq^+(x_j^--x_4^-)}\nn\\
  && \int\frac{dk_1^+}{2\pi}e^{-ik_1^+(x_5^--x_i^-)}
  \int\frac{d(k-k_1)^+}{2\pi}e^{-i(k-k_1)^+(x_5-x_j)^-}\nn\\
&& \int \frac{dl^-}{2\pi}e^{-il^-(x_3^+-x_1^+)}  \int\frac{d(k_2-l)^-}{2\pi}e^{-i(k_2-l)^-(x_3^+-x_2^+)}
   \int \frac{dk_2^-}{2\pi}e^{-ik_2^-(x_i^+-x_3^+)}  \int \frac{dq^-}{2\pi}e^{-iq^-(x_j^+-x_4^+)}\nn\\
      && \int \frac{dk_1^-}{2\pi}e^{-ik_1^-(x_5^+-x_i^+)}
      \int\frac{d(k-k_1)^-}{2\pi}e^{-i(k-k_1)^-(x_5^+-x_j^+)} \nn\\
&&\int\frac{d^2l_{\perp}}{(2\pi)^2}e^{il_{\perp}({x_3}_\perp-x_{1\perp})}
  \int\frac{d^2(k_2-l)_\perp}{(2\pi)^2}e^{i(k_2-l)_\perp.(x_{3\perp}-x_{2\perp})}
  \int\frac{d^2k_{2\perp}}{(2\pi)^2}e^{ilk_{2\perp}({x_i}_\perp-x_{3\perp})}
   \int\frac{d^2q_{\perp}}{(2\pi)^2}e^{iq_{\perp}({x_j}_\perp-x_{4\perp})}\nn\\
   &&\int\frac{d^2k_{1\perp}}{(2\pi)^2}e^{ik_{1\perp}({x_5}_\perp-x_{i\perp})}
   \int\frac{d^2(k-k_1)_\perp}{(2\pi)^2}e^{i(k-k_1)_\perp.(x_{5\perp}-x_{j\perp})}\nn\\
   &&\frac{-i}{2l^+l^--l_{\perp}^2+i\epsilon l^+}~\frac{-i}{2(k_2^+-l^+)(k_2-l)^--(k_2-l)_{\perp}^2+i\epsilon(k_2-l)^+}
  ~\frac{-i}{2(k-k_1)^+q^--q_{\perp}^2+i\epsilon (k-k_1)^+}~\nn\\ 
  &&\frac{-i}{2k_2^+k_2^--k_{2\perp}^2+i\epsilon k_2^+}~\frac{-i}{2k_1^+k_1^--k_{1\perp}^2+i\epsilon k_1^+}~\frac{-i}{2(k-k_1)^+(k-k_1)^--(k-k_1)_{\perp}^2+i\epsilon(k-k_1)^+}\nn\\
&&(-gf^{srt})(-gf^{bac})\sum_{\lambda_1,\lambda_2,\lambda_3,\lambda_4=\pm1}
\left[(2l-k_2).\epsilon_{\lambda_1}^*(k_2)\epsilon_{\lambda_3}.\epsilon_{\lambda_2}
 -(k_2+l).\epsilon_{\lambda_4}(k_2-l)\epsilon_{\lambda_3}.\epsilon_{\lambda_1}^*
 +(2k_2-l).\epsilon_{\lambda_3}(l)\epsilon_{\lambda_1}^*.\epsilon_{\lambda_2}\right]\nn\\
 &&\left[(2k_1-k).\epsilon_{\lambda}^*(k)\epsilon_{\lambda_2}.\epsilon_{\lambda_1}
 -(k+k_1).\epsilon_{\lambda_2}(k-k_1)\epsilon_{\lambda_1}.\epsilon_{\lambda}^*
 +(2k-k_1).\epsilon_{\lambda_1}(k_1)\epsilon_{\lambda}^*.\epsilon_{\lambda_2}\right]
 \epsilon_{\lambda_3}^{*-}(l)\epsilon_{\lambda_4}^{*-}(k_2-l)\epsilon_{\lambda_2}^{*-}(q)\label{SA1}
\end{eqnarray}

\vspace{0.5cm}
Now integrating Eq.\eqref{SA1} over light cone time, $x_1^+$, $x_2^+$, $x_3^+$, $x_4^+$ and $x_5^+$, and then over $'-'$ component of momentum we get,
       
 
\begin{eqnarray}
A_1&=&(ig)^3(-gf^{srt})(-gf^{bac})(-i)\int\frac{d^2l_{\perp}}{(2\pi)^2}e^{il_{\perp}({x_3}_\perp-x_{1\perp})}
  \int\frac{d^2(k_2-l)_\perp}{(2\pi)^2}e^{i(k_2-l)_\perp.(x_{3\perp}-x_{2\perp})}
  \int\frac{d^2k_{2\perp}}{(2\pi)^2}e^{ilk_{2\perp}({x_i}_\perp-x_{3\perp})}\nn\\
   &&\int\frac{d^2q_{\perp}}{(2\pi)^2}e^{iq_{\perp}({x_j}_\perp-x_{4\perp})}\int\frac{d^2k_{1\perp}}{(2\pi)^2}e^{ik_{1\perp}({x_5}_\perp-x_{i\perp})}
   \int\frac{d^2(k-k_1)_\perp}{(2\pi)^2}e^{i(k-k_1)_\perp.(x_{5\perp}-x_{j\perp})}\nn\\
&&(2k_1^+)U^{br}_{x_{5\perp}}~\left(2(k-k_1)^+\right)U^{cd}_{x_{4\perp}} (V_{x_{1\perp}}t^s)_1(V_{x_{2\perp}}t^dt^t)_2\nn \\
&& \int\frac{dl^+}{2\pi}e^{-il^+(x_3^--x_1^-)}   \int\frac{d(k_2-l)^+}{2\pi}e^{-i(k_2-l)^+(x_3-x_2)^-}
 \int\frac{dk_2^+}{2\pi}e^{-ik_2^+(x_i^--x_3^-)}  \int\frac{dq^+}{2\pi}e^{-iq^+(x_j^--x_4^-)}\nn\\
  &&\int\frac{dk_1^+}{2\pi}e^{-ik_1^+(x_5^--x_i^-)}\int\frac{d(k-k_1)^+}{2\pi}e^{-i(k-k_1)^+(x_5-x_j)^-}
  \nn\\
  &&\frac{-i}{l_{\perp}^2 -3i\epsilon l^+}
   \frac{-i}{(k_2-l)_{\perp}^2-3i\epsilon(k_2-l)^+}~
   \frac{-i}{k_{2\perp}^2-5i\epsilon k_2^+}\frac{-i}{q_{\perp}^2 -3i\epsilon (k-k_1)^+}\frac{1}{2k^+(k_{1\perp}-\alpha_2k_\perp)^2}\nn\\
  &&\sum_{\lambda_1,\lambda_2,\lambda_3,\lambda_4=\pm1}
\left[(2l-k_2).\epsilon_{\lambda_1}^*(k_2)\epsilon_{\lambda_3}.\epsilon_{\lambda_2}
 -(k_2+l).\epsilon_{\lambda_4}(k_2-l)\epsilon_{\lambda_3}.\epsilon_{\lambda_1}^*
 +(2k_2-l).\epsilon_{\lambda_3}(l)\epsilon_{\lambda_1}^*.\epsilon_{\lambda_2}\right]\nn\\
 &&\left[(2k_1-k).\epsilon_{\lambda}^*(k)\epsilon_{\lambda_2}.\epsilon_{\lambda_1}
 -(k+k_1).\epsilon_{\lambda_2}(k-k_1)\epsilon_{\lambda_1}.\epsilon_{\lambda}^*
 +(2k-k_1).\epsilon_{\lambda_1}(k_1)\epsilon_{\lambda}^*.\epsilon_{\lambda_2}\right]
 \epsilon_{\lambda_3}^{*-}(l)\epsilon_{\lambda_4}^{*-}(k_2-l)\epsilon_{\lambda_2}^{*-}(q)\label{SA2}
\end{eqnarray}
%
  \vspace{.5cm}
 In the Eq.\eqref{SA2} we see that
 integration over all the $+$-component of momenta can be reduced in terms of only independent momentum, $i.e.$\\
 \begin{eqnarray}
 && \int\frac{dl^+}{2\pi}e^{-il^+(x_3^--x_1^-)}   \int\frac{d(k_2-l)^+}{2\pi}e^{-i(k_2-l)^+(x_3-x_2)^-}
 \int\frac{dk_2^+}{2\pi}e^{-ik_2^+(x_i^--x_3^-)}  \int\frac{dq^+}{2\pi}e^{-iq^+(x_j^--x_4^-)}\nn\\
  &&\int\frac{dk_1^+}{2\pi}e^{-ik_1^+(x_5^--x_i^-)}\int\frac{d(k-k_1)^+}{2\pi}e^{-i(k-k_1)^+(x_5-x_j)^-}\nn\\
  &&\equiv \int\frac{dl^+}{2\pi}e^{-il^+(x_2^--x_1^-)}  \int\frac{dk_1^+}{2\pi}e^{-ik_1^+(x_4^--x_2^-)}~,
 \end{eqnarray}

\vspace{0.5cm}
We now define the variables $\alpha_1=l^+/k^+$  and $\alpha_2=k_1^+/k^+$.
Next we have to perform the polarization sum over four different polarization vector, $\epsilon_{\lambda_i}(i=1,2,3,4)$. Explicit calculation of polarization sum is described in Appendix \ref{PS}. After this Eq.\eqref{SA2} reduces to, 

 \begin{eqnarray}
A_1&=&(ig)^3(-gf^{srt})(-gf^{bac})\int\frac{d^2l_{\perp}}{(2\pi)^2}e^{il_{\perp}({x_3}_\perp-x_{1\perp})}
  \int\frac{d^2(k_2-l)_\perp}{(2\pi)^2}e^{i(k_2-l)_\perp.(x_{3\perp}-x_{2\perp})}
  \int\frac{d^2k_{2\perp}}{(2\pi)^2}e^{ilk_{2\perp}({x_i}_\perp-x_{3\perp})}\nn\\
   &&\int\frac{d^2q_{\perp}}{(2\pi)^2}e^{iq_{\perp}({x_j}_\perp-x_{4\perp})}\int\frac{d^2k_{1\perp}}{(2\pi)^2}e^{ik_{1\perp}({x_5}_\perp-x_{i\perp})}
   \int\frac{d^2(k-k_1)_\perp}{(2\pi)^2}e^{i(k-k_1)_\perp.(x_{5\perp}-x_{j\perp})}\nn\\
&&U^{br}_{x_{5\perp}}~U^{cd}_{x_{4\perp}} (V_{x_{1\perp}}t^s)_1(V_{x_{2\perp}}t^dt^t)_2~~\frac{1}{q_\perp^2~l_\perp^2~(k_2-l)_\perp^2~k_{2\perp}^2}\nn \\
&& \int\frac{d\alpha_1}{2\pi}e^{-ik^+\alpha_1(x_2^--x_1^-)}  \int\frac{d\alpha_2}{2\pi}e^{-ik^+\alpha_2(x_4^--x_2^-)}\left[\frac{2\alpha_2}{\alpha_1(\alpha_2-\alpha_1)}\frac{1}{(k_{1\perp}-\alpha_2k_\perp)^2}\right]\nn\\
&&4\left[\frac{l_\perp.(k_2-l)_\perp}{\alpha_2}\left(q_\perp.(\alpha_2l_\perp-\alpha_1k_{2\perp})(k_{1\perp}-\alpha_2k_\perp).\epsilon_{\lambda_\perp}^*~-\frac{1}{(1-\alpha_2)}q_\perp.(k_{1\perp}-\alpha_2k_\perp)~(\alpha_2l_\perp-\alpha_1k_{2\perp}).\epsilon_{\lambda_\perp}^*\right.\right.\nn\\
&&~~~~~~~~~~~~~~~~~~~~~~~~~~~~~~~\left.\left.-\frac{1}{\alpha_2}(k_{1\perp}-\alpha_2k_\perp).(\alpha_2l_\perp-\alpha_1k_{2\perp})~q_\perp.\epsilon_{\lambda_\perp}^*\right)\right.\nn\\
&&\left.-\frac{(\alpha_2l_\perp-\alpha_1k_{2\perp}).(k_2-l)_\perp}{(\alpha_2-\alpha_1)}\left((k_{1\perp}-\alpha_2k_\perp).\epsilon_{\lambda_\perp}^*~l_\perp.q_\perp-\frac{1}{1-\alpha_2}(k_{1\perp}-\alpha_2k_\perp).q_\perp~l_\perp.\epsilon_{\lambda_\perp}^*\right.\right.\nn\\
&&~~~~~~~~~~~~~~~~~~~~~~~~~~~~~~~\left.\left.-\frac{1}{\alpha_2}(k_{1\perp}-\alpha_2k_\perp).l_\perp~q_\perp.\epsilon_{\lambda_\perp}^*\right)\right.\nn\\
&&\left.-\frac{(\alpha_2l_\perp-\alpha_1k_{2\perp}).l_\perp}{\alpha_1}\left((k_{1\perp}-\alpha_2k_\perp).\epsilon_{\lambda_\perp}^*~(k_2-l)_\perp.q_\perp-\frac{1}{1-\alpha_2}(k_{1\perp}-\alpha_2k_\perp).q_\perp~(k_2-l)_\perp.\epsilon_{\lambda_\perp}^*\right.\right.\nn\\
&&~~~~~~~~~~~~~~~~~~~~~~~~~~~~~~~\left.\left.-\frac{1}{\alpha_2}(k_{1\perp}-\alpha_2k_\perp).(k_2-l)_\perp~q_\perp.\epsilon_{\lambda_\perp}^*\right)\right]~.
\end{eqnarray}

 \vspace{0.5cm}

 \section{Final expression in transverse momentum space}
Finally we need to perform integration over $+$-component of momentum.
 %
 We redefine variables $\alpha_1$ and $\alpha_2$,
\begin{eqnarray}
       \alpha&\equiv&\alpha_2\nn\\
       \beta&\equiv&\frac{\alpha_1}{\alpha_2}\nn\\
       \int  d\alpha_1 \int d\alpha_2&\equiv& -\int \alpha  d\alpha \int d\beta\nn
\end{eqnarray}
 so that the integration becomes,

\begin{eqnarray}
&& \int\frac{d\alpha}{2\pi} \int\frac{d\beta}{2\pi}e^{-ik^+\alpha\beta(x_2^--x_1^-)} \left[\frac{-2\alpha}{\alpha}{\rm{PV}}\left(\frac{1}{\beta}\right){\rm{PV}}\left(\frac{1}{1-\beta}\right)\frac{1}{(k_{1\perp}-\alpha k_\perp)^2}\right]\nn\\
&&4\left[{l_\perp.(k_2-l)_\perp}\left(q_\perp.(l_\perp-\beta k_{2\perp})(k_{1\perp}-\alpha k_\perp).\epsilon_{\lambda_\perp}^*~-{\rm{PV}}\left(\frac{1}{(1-\alpha)}\right)q_\perp.(k_{1\perp}-\alpha k_\perp)~(l_\perp-\beta k_{2\perp}).\epsilon_{\lambda_\perp}^*\right.\right.\nn\\
&&~~~~~~~~~~~~~~~~~~~~~~~~~~~~~~~\left.\left.-{\rm{PV}}\left(\frac{1}{\alpha}\right)(k_{1\perp}-\alpha k_\perp).(l_\perp-\beta k_{2\perp})~q_\perp.\epsilon_{\lambda_\perp}^*\right)\right.\nn\\
&&\left.-{(l_\perp-\beta k_{2\perp}).(k_2-l)_\perp}{\rm{PV}}\left(\frac{1}{1-\beta}\right)\left((k_{1\perp}-\alpha k_\perp).\epsilon_{\lambda_\perp}^*~l_\perp.q_\perp-{\rm{PV}}\left(\frac{1}{(1-\alpha)}\right)(k_{1\perp}-\alpha k_\perp).q_\perp~l_\perp.\epsilon_{\lambda_\perp}^*\right.\right.\nn\\
&&~~~~~~~~~~~~~~~~~~~~~~~~~~~~~~~\left.\left.-{\rm{PV}}\left(\frac{1}{\alpha}\right)(k_{1\perp}-\alpha k_\perp).l_\perp~q_\perp.\epsilon_{\lambda_\perp}^*\right)\right.\nn\\
&&\left.-{(l_\perp-\beta k_{2\perp}).l_\perp}{\rm{PV}}\left(\frac{1}{\beta}\right)\left((k_{1\perp}-\alpha k_\perp).\epsilon_{\lambda_\perp}^*~(k_2-l)_\perp.q_\perp-{\rm{PV}}\left(\frac{1}{(1-\alpha)}\right)(k_{1\perp}-\alpha k_\perp).q_\perp~(k_2-l)_\perp.\epsilon_{\lambda_\perp}^*\right.\right.\nn\\
&&~~~~~~~~~~~~~~~~~~~~~~~~~~~~~~~\left.\left.-{\rm{PV}}\left(\frac{1}{\alpha}\right)(k_{1\perp}-\alpha k_\perp).(k_2-l)_\perp~q_\perp.\epsilon_{\lambda_\perp}^*\right)\right] . \label{di}   
\end{eqnarray}

\vspace{0.5cm}

An important observation at this stage of amplitude calculation of ${\cal{O}}(g^5)$ is that, while there are two momentum integral, there is only one exponential factor, $e^{-ik^+\alpha\beta(x_2^--x_1^-)}$, with $(x_2^--x_1^-)$. The presence of $(x_2^--x_1^-)$ in the exponential makes the integral convergent otherwise it is divergent. We follow straightforward procedure using residues to solve $\beta-$ integral (Appendix \ref{beta}) to get, 

\begin{eqnarray}
  &&\int\frac{d\alpha}{2\pi}  \left[\frac{1}{(k_{1\perp}-\alpha k_\perp)^2}\right]\frac{i}{2}~{\rm{Sign}}(x_2^--x_1^-)\nn\\  
  &&\left[(k_{1\perp}-\alpha k_\perp).\epsilon_{\lambda_\perp}^*\left\{{l_\perp.(k_2-l)_\perp}\left(q_\perp.l_\perp-q_\perp.(l_\perp-k_{2\perp})e^{-ik^+\alpha(x_2^--x_1^-)}\right)\right.\right.\nn\\
 &&~~~~~~~~~~~~~~~~~~~~~~~~~~~\left.\left. -\left(( k_{2\perp}-l_{\perp}).l_\perp-(k_{2\perp}-l_{\perp}).l_\perp e^{-ik^+\alpha(x_2^--x_1^-)}\right)l_\perp.q_\perp\right.\right.\nn\\
  &&~~~~~~~~~~~~~~~~~~~~~~~~~\left.\left.-\left((l_{\perp}- k_{2\perp}).l_\perp-(l_{\perp}- k_{2\perp}).l_\perp e^{-ik^+\alpha(x_2^--x_1^-)}\right)(k_2-l)_\perp.q_\perp\right\}\right.\nn\\
  &&\left.-{\rm{PV}}\left(\frac{1}{(1-\alpha)}\right)q_\perp.(k_{1\perp}-\alpha k_\perp)\left\{{l_\perp.(k_2-l)_\perp}~\left(l_\perp.\epsilon_{\lambda_\perp}^*-(l_\perp-k_{2\perp}).\epsilon_{\lambda_\perp}^*e^{-ik^+\alpha(x_2^--x_1^-)}\right)\right.\right.\nn\\
 &&~~~~~~~~~~~~~~~~~~\left.\left. -l_\perp.\epsilon_{\lambda_\perp}^*\left(( k_{2\perp}-l_{\perp}).l_\perp-(k_{2\perp}-l_{\perp}).l_\perp e^{-ik^+\alpha(x_2^--x_1^-)}\right)\right.\right.\nn\\
 &&~~~~~~~~~~~~~~~~~~~~~~~~\left.\left.-(k_2-l)_\perp.\epsilon_{\lambda_\perp}^*\left((l_{\perp}- k_{2\perp}).l_\perp-(l_{\perp}- k_{2\perp}).l_\perp e^{-ik^+\alpha(x_2^--x_1^-)}\right)
 \right\}\right.\nn\\
 &&\left.-{\rm{PV}}\left(\frac{1}{\alpha}\right)q_\perp.\epsilon_{\lambda_\perp}^*\left\{{l_\perp.(k_2-l)_\perp}\left((k_{1\perp}-\alpha k_\perp).l_\perp-(k_{1\perp}-\alpha k_\perp).(l_\perp-k_{2\perp})e^{-ik^+\alpha(x_2^--x_1^-)}\right)~\right.\right.\nn\\
 &&~~~~~~~~~~~~~~~~~~~~~~~~~\left.\left.-\left(( k_{2\perp}-l_{\perp}).l_\perp-(k_{2\perp}-l_{\perp}).l_\perp e^{-ik^+\alpha(x_2^--x_1^-)}\right)(k_{1\perp}-\alpha k_\perp).l_\perp\right.\right.\nn\\
 &&~~~~~~~~~~~~~~~~~~~~~~~~~~~~~\left.\left.-\left((l_{\perp}- k_{2\perp}).l_\perp-(l_{\perp}- k_{2\perp}).l_\perp e^{-ik^+\alpha(x_2^--x_1^-)}\right)(k_{1\perp}-\alpha k_\perp).(k_2-l)_\perp \right\}
  \right]
\end{eqnarray}

We use the following to evaluate the diverging part,
\begin{eqnarray}
       \oint f(z)dz&=&\lim_{R\rightarrow\infty}\int_{-R}^{R}f(x)dx+\lim_{R\rightarrow\infty}\int_{C}f(z)dz\nn\\
       &&~~~~~~~~~~~~~~~{\rm{or}}\nn\\
       \lim_{R\rightarrow\infty}\int_{-R}^{R}f(x)dx &=&2\pi i\sum {\rm{residues~(upper~ half-plane)}}-\lim_{R\rightarrow\infty}\int_{C}f(z)dz
\end{eqnarray}
For one of the terms in present case we have,
\begin{eqnarray}
       \lim_{R\rightarrow\infty}\int_{C}f(z)dz&=&\lim_{|\alpha|\rightarrow\infty}\int_{C}\frac{d\alpha}{2\pi}\left[\frac{(k_{1\perp}-\alpha k_\perp).\epsilon_{\lambda_\perp}^*}{(k_{1\perp}-\alpha k_\perp)^2}\right]\nn\\
       &=&\int_{C}\frac{d\alpha}{2\pi}\lim_{|\alpha|\rightarrow\infty}\left[\frac{(k_{1\perp}-\alpha k_\perp).\epsilon_{\lambda_\perp}^*}{(k_{1\perp}-\alpha k_\perp)^2}\right]\nn\\
       &=&\int_{C}\frac{d\alpha}{2\pi}\left[\frac{-\alpha k_\perp.\epsilon_{\lambda_\perp}^*}{\alpha^2 k_\perp^2}\right]  ~ ~ ~ . . . ~ ~ ~ (0\leq {\rm{arg}}~\alpha\leq\pi)\nn\\
       &=&-\frac{i\epsilon_{\lambda_\perp}^*.k_\perp}{2k_\perp^2}~.
\end{eqnarray}
and, 
\begin{eqnarray}
       2\pi i\sum {\rm{residues~(upper~ half-plane)}}=\frac{[k_\perp^2k_{1\perp}-k_\perp(k_{1\perp}.k_\perp)].\epsilon_{\lambda\perp}^*}{2k_\perp^2|k_{1\perp}\times k_\perp|}-i\frac{\epsilon_{\lambda_\perp}^*.k_\perp}{2k_\perp^2}~.
\end{eqnarray}
%
Using above we solve each term, shown in Appendix \ref{alpha} and the remaining converging part (also in Appendix \ref{alpha}) is calculated using residues. Combined together the  amplitude for the sample diagram in transverse momentum space is, 
  \begin{empheq}{align}
  A_1=&~ 4g^5f^{srt}f^{bac}\int\frac{d^2l_{\perp}}{(2\pi)^2}e^{il_{\perp}({x_3}_\perp-x_{1\perp})}
  \int\frac{d^2(k_2-l)_\perp}{(2\pi)^2}e^{i(k_2-l)_\perp.(x_{3\perp}-x_{2\perp})}
  \int\frac{d^2k_{2\perp}}{(2\pi)^2}e^{ilk_{2\perp}({x_i}_\perp-x_{3\perp})}\nn\\
   &\int\frac{d^2q_{\perp}}{(2\pi)^2}e^{iq_{\perp}({x_j}_\perp-x_{4\perp})}\int\frac{d^2k_{1\perp}}{(2\pi)^2}e^{ik_{1\perp}({x_5}_\perp-x_{i\perp})}
   \int\frac{d^2(k-k_1)_\perp}{(2\pi)^2}e^{i(k-k_1)_\perp.(x_{5\perp}-x_{j\perp})}\nn\\
&U^{br}_{x_{5\perp}}~U^{cd}_{x_{4\perp}} (V_{x_{1\perp}}t^s)_1(V_{x_{2\perp}}t^dt^t)_2~~\frac{1}{q_\perp^2~l_\perp^2~(k_2-l)_\perp^2~k_{2\perp}^2}~~{l_\perp.(k_2-l)_\perp}\nn \\
&  \left[\left\{k_{2\perp}.q_\perp\frac{\epsilon_{\lambda_\perp}^*\times k_\perp}{k_{\perp}^2}+k_{2\perp}.\epsilon_{\lambda_\perp}^*\frac{q_\perp\times(k_\perp-k_{1\perp})}{(k_\perp-k_{1\perp})^2}\right.\right.\nn\\
   &~~~~~~~~~~~~~~~~~~~~~~~~~~~~~~~~~~~~~~~\left.\left.+2q_\perp.\epsilon_{\lambda_\perp}^*\frac{(k_2-l)_\perp\times k_{1\perp}}{k_{1\perp}^2} \right\}{\rm{Sign}}(x_2^--x_1^-)~{\rm{Sign}}(k_\perp\times k_{1\perp})-i\frac{l_\perp.q_\perp}{k_\perp^2}\right].   
 \end{empheq}

 \section{outlook}       
To find the order $\alpha_s^3$ contribution to the classical gluon production cross section one needs to calculate all diagrams of ${\cal O}(g^6)$. Diagrams can be separated into three classes: (i) the square of the order-$g^3$ amplitude; (ii) the interference between the order-$g^5$ amplitude and the leading-order (order-$g$)  amplitude (iii) the interference between the order-$g^4$ amplitude and the order-$g^2$ amplitude. The fact that all the propagators are retarted implies that only ${\cal O}(g^3)$ and ${\cal O}(g^5)$ will give complete cross section. All ${\cal{O}}(g^3)$ amplitudes are calculated in \cite{Chirilli:2015tea}. In this article we report the calculation, in full details, of one of the sample diagram from the set of diagrams that contribute in ${\cal O}(g^5)$. Complete calculations of all would be reported elsewhere in the future.

 \begin{acknowledgments}
 I thank Raktim Abir for insightful discussions and advice in the development of this work. I am grateful to Mariyah Siddiqah and Khatiza Banu for their assistance and support during the various stages of this work. 
 \end{acknowledgments}

\appendix
\section{Calculating numerator}\label{Num}

\subsubsection*{Solving Numerator of propagators (without instantaneous contribution) and 3-gluon vertices}
\noindent $\bullet$ $D^{\alpha\alpha^'}(l)=\sum_{\lambda_3}\epsilon_{\lambda_3}^{*\alpha}(l)\epsilon_{\lambda_3}^{\alpha^'}(l)$
~~~~~~
\noindent $\bullet$ $D^{\beta\beta^'}(k_2-l)=\sum_{\lambda_4}\epsilon_{\lambda_4}^{*\beta}(k_2-l)\epsilon_{\lambda_3}^{\beta^'}
             (k_2-l)$\\
\noindent $\bullet$ $D_{\mu}^{\mu^'(k_2)}=\sum_{\lambda_1^'}\epsilon_{\lambda_1^'}^{*\mu^'}(k_2)\epsilon_{\lambda_1^'}^{\mu^'}
            (k_2)$~~~
\noindent $\bullet$ $D^{\mu^'\rho}(k_1)=\sum_{\lambda_1}\epsilon_{\lambda_1}^{*\mu^'}(k_1)\epsilon_{\lambda_1}^{\rho}(k_1)$\\
\noindent $\bullet$ $D^{\gamma\gamma^'}(q)=\sum_{\lambda_2^'}\epsilon_{\lambda_2^'}^{*\gamma}(q)\epsilon_{\lambda_2^'}^{\gamma^'}
            (q)$~~~~~~
\noindent $\bullet$ $D_{\gamma^'}^{\gamma}(k-k_1)=\sum_{\lambda_2}\epsilon_{\lambda_2\gamma^'}^*(k_2)
           \epsilon_{\lambda_2}^{\sigma}(k-k_1)$
\\~\\
Multiplying 3-gluon vertex at $\rm{X_3}$ and propagators with momenta $l, (k_2-l), k_2, k_1$. 
\begin{eqnarray}
 &&-gf^{srt}\left[(2l-k_2)_{\mu}g_{\alpha^'\beta^'}-(k_2+l)_{\beta^'}g_{\alpha^'\mu}+(2k_2-l)_{\alpha^'}g_{\mu\beta^'}\right]\nn\\
 &&\times \epsilon_{\lambda_3}^{*\alpha}(l)\epsilon_{\lambda_3}^{\alpha^'}(l)~
 \epsilon_{\lambda_4}^{*\beta}(k_2-l)\epsilon_{\lambda_3}^{\beta^'}(k_2-l)
 ~\epsilon_{\lambda_1^'}^{*\mu^'}(k_2)\epsilon_{\lambda_1^'}^{\mu^'}(k_2)
 ~\epsilon_{\lambda_1}^{*\mu^'}(k_1)\epsilon_{\lambda_1}^{\rho}(k_1)
\end{eqnarray}
using,~$\epsilon_{\lambda_1^'}^{\mu^'}(k_2)\epsilon_{\lambda_1}^{*\mu^'}(k_1)=\delta_{\lambda_1\lambda_1^'}$ and contracting
above expression.
\begin{eqnarray}
 =-gf^{srt}\left[(2l-k_2).\epsilon_{\lambda_1}^*(k_2)\epsilon_{\lambda_3}.\epsilon_{\lambda_2}
 -(k_2+l).\epsilon_{\lambda_4}(k_2-l)\epsilon_{\lambda_3}.\epsilon_{\lambda_1}^*
 +(2k_2-l).\epsilon_{\lambda_3}(l)\epsilon_{\lambda_1}^*.\epsilon_{\lambda_2}\right]\epsilon_{\lambda_3}^{*\alpha}(l)
 \epsilon_{\lambda_4}^{*\beta}(k_2-l)\epsilon_{\lambda_1}^{\rho}(k_1)
\end{eqnarray}

Multiplying 3-gluon vertex at $\rm{X_5}$ and propagators with momenta $q, (k-k_1)$ to eq(8).  
\begin{eqnarray}
&&-gf^{bac}\left[(2k_1-k)_{\nu}g_{\rho\sigma}-(k+k_1)_{\sigma}g_{\rho\nu}+(2k-k_1)_{\rho}g_{\nu\sigma}\right]\epsilon_\lambda^{\nu*}(k)
\times \epsilon_{\lambda_2\gamma^'}^*(k_2)\epsilon_{\lambda_2}^{\sigma}(k-k_1) 
\epsilon_{\lambda_2^'}^{*\gamma}(q)\epsilon_{\lambda_2^'}^{\gamma^'}(q)\nn\\
&&\times (-gf)^{srt}\left[(2l-k_2).\epsilon_{\lambda_1}^*(k_2)\epsilon_{\lambda_3}.\epsilon_{\lambda_2}
 -(k_2+l).\epsilon_{\lambda_4}(k_2-l)\epsilon_{\lambda_3}.\epsilon_{\lambda_1}^*
 +(2k_2-l).\epsilon_{\lambda_3}(l)\epsilon_{\lambda_1}^*.\epsilon_{\lambda_2}\right]\epsilon_{\lambda_3}^{*\alpha}(l)
 \epsilon_{\lambda_4}^{*\beta}(k_2-l)\epsilon_{\lambda_1}^{\rho}(k_1)\nn
\end{eqnarray}

Again using,~$\epsilon_{\lambda_2^'}^{\gamma^'}(q)\epsilon_{\lambda_2\gamma^'}^{*}(k-k_1)=\delta_{\lambda_2\lambda_2^'}$ 
and contracting above expression.
\begin{eqnarray}
 &=&(-gf^{srt})(-gf^{bac})\left[(2l-k_2).\epsilon_{\lambda_1}^*(k_2)\epsilon_{\lambda_3}.\epsilon_{\lambda_2}
 -(k_2+l).\epsilon_{\lambda_4}(k_2-l)\epsilon_{\lambda_3}.\epsilon_{\lambda_1}^*
 +(2k_2-l).\epsilon_{\lambda_3}(l)\epsilon_{\lambda_1}^*.\epsilon_{\lambda_2}\right]\nn\\
 &&\left[(2k_1-k).\epsilon_{\lambda}^*(k)\epsilon_{\lambda_2}.\epsilon_{\lambda_1}
 -(k+k_1).\epsilon_{\lambda_2}(k-k_1)\epsilon_{\lambda_1}.\epsilon_{\lambda}^*
 +(2k-k_1).\epsilon_{\lambda_1}(k_1)\epsilon_{\lambda}^*.\epsilon_{\lambda_2}\right]
 \epsilon_{\lambda_3}^{*\alpha}(l)\epsilon_{\lambda_4}^{*\beta}(k_2-l)\epsilon_{\lambda_2^'}^{*\gamma}(q)\label{A3}
\end{eqnarray}
In \eqref{A3} we choose $'-'$-component of $\alpha, \beta$ and $\gamma$ indices as only this gives leading contribution.

   \section{Polarization Sum}\label{PS}
 \begin{eqnarray}
 &&\sum_{\lambda_1,\lambda_2,\lambda_3,\lambda_4=\pm1} \epsilon_{\lambda_{3\perp}}^{*}.l_\perp~\epsilon_{\lambda_{4\perp}}^{*}.(k_2-l)_\perp~
\epsilon_{\lambda_{2\perp}}^{*}.q_\perp\nn\\
&&\left[(2l-k_2).\epsilon_{\lambda_1}^*(k_2)\epsilon_{\lambda_3}.\epsilon_{\lambda_2}
 -(k_2+l).\epsilon_{\lambda_4}(k_2-l)\epsilon_{\lambda_3}.\epsilon_{\lambda_1}^*
 +(2k_2-l).\epsilon_{\lambda_3}(l)\epsilon_{\lambda_1}^*.\epsilon_{\lambda_2}\right]\nn\\
 &&\left[(2k_1-k).\epsilon_{\lambda}^*(k)\epsilon_{\lambda_2}.\epsilon_{\lambda_1}
 -(k+k_1).\epsilon_{\lambda_2}(k-k_1)\epsilon_{\lambda_1}.\epsilon_{\lambda}^*
 +(2k-k_1).\epsilon_{\lambda_1}(k_1)\epsilon_{\lambda}^*.\epsilon_{\lambda_2}\right]
\end{eqnarray}
 Simplifying the dot product and writing above in terms of $\alpha_1$ and $\alpha_2$.\\
  \begin{eqnarray}
 &&\sum_{\lambda_1,\lambda_2,\lambda_3,\lambda_4=\pm1} \epsilon_{\lambda_{3\perp}}^{*}.l_\perp~\epsilon_{\lambda_{4\perp}}^{*}.(k_2-l)_\perp~
\epsilon_{\lambda_{2\perp}}^{*}.q_\perp\nn\\
&&\left[-2(\alpha_2l_\perp-\alpha_1k_{2\perp}).\epsilon_{\lambda_{1\perp}}^*\delta_{\lambda_3-\lambda_4}
 -\frac{2}{\alpha_2-\alpha_1}(\alpha_2l_\perp-\alpha_1k_{2\perp}).\epsilon_{\lambda_{4\perp}}\delta_{\lambda_1\lambda_4}
 -\frac{2}{\alpha_1}(\alpha_2l_\perp-\alpha_1k_{2\perp}).\epsilon_{\lambda_{3\perp}}\delta_{\lambda_3\lambda_1}\right]\nn\\
 &&\left[-2(k_{1\perp}-\alpha k_\perp).\epsilon_{\lambda_\perp}^*\delta_{\lambda_2-\lambda_1}
 -\frac{2}{1-\alpha_2}(k_{1\perp}-\alpha k_\perp).\epsilon_{\lambda_{2\perp}}\delta_{\lambda\lambda_1}
 -\frac{2}{\alpha_2}(k_{1\perp}-\alpha_2 k_\perp).\epsilon_{\lambda_{1\perp}}\delta_{\lambda\lambda_2}\right]
\end{eqnarray}
where, we have used:\\~\\
        $\epsilon_{\lambda_1\perp}\epsilon_{\lambda_2\perp}^*=1/2(\lambda_1\lambda_2-1)
        \Rightarrow\lambda_1=-\lambda_2$ or $\delta_{\lambda_1-\lambda_2}$
 and $\epsilon_{\lambda_1\perp}\epsilon_{\lambda_2\perp}=1/2(\lambda_1\lambda_2+1)
        \Rightarrow\lambda_1=\lambda_2$ or $\delta_{\lambda_1\lambda_2}$\\
Consider one term solved as,
\begin{eqnarray}
 &&4\sum_{\lambda_1,\lambda_2,\lambda_3,\lambda_4=\pm1} \epsilon_{\lambda_{3\perp}}^{*}.l_\perp~\epsilon_{\lambda_{4\perp}}^{*}.(k_2-l)_\perp~
\epsilon_{\lambda_{2\perp}}^{*}.q_\perp  \left[(\alpha_2l_\perp-\alpha_1k_{2\perp}).\epsilon_{\lambda_{1\perp}}^*\delta_{\lambda_3-\lambda_4} \right]\left[-(k_{1\perp}-\alpha_2 k_\perp).\epsilon_{\lambda_\perp}^*\delta_{\lambda_2-\lambda_1}\right]\nn\\
&=&4(k_{1\perp}-\alpha_2 k_\perp).\epsilon_{\lambda_\perp}^*\sum_{\lambda_1}(\alpha_2l_\perp-\alpha_1k_{2\perp}).\epsilon_{\lambda_{1\perp}}^*\epsilon_{\lambda_{-1\perp}}^{*}.q_\perp\sum_{\lambda_3}\epsilon_{\lambda_{4\perp}}^{*}.(k_2-l)_\perp~\epsilon_{\lambda_{3\perp}}^{*}.l_\perp\nn\\
&=&4(k_{1\perp}-\alpha_2 k_\perp).\epsilon_{\lambda_\perp}^*\frac{1}{2}\sum_{\lambda_1}\left\{(\alpha_2l-\alpha_1k_{2})_x,(\alpha_2l-\alpha_1k_{2})_y\right\}\left\{\lambda_1,-i\right\}\left\{q_x,q_y\right\}\left\{-\lambda_1,-i\right\}\nn\\
&&~~~~~~~~~~~~~~~~~~~~~~~\frac{1}{2}\sum_{\lambda_3}\left\{(k_2-l)_x,(k_2-l)_y\right\}\left\{-\lambda_3,-i\right\}\left\{l_x,l_y\right\}\left\{\lambda_3,-i\right\}\nn\\
&=&4(k_{1\perp}-\alpha_2 k_\perp).\epsilon_{\lambda_\perp}^*\left[(\alpha_2l_\perp-\alpha_1k_{2\perp}).q_\perp\right]~\left[(k_{2\perp}-l_\perp).l_\perp\right]
\end{eqnarray}
similarly we calculate all 9 terms.
\section{$\beta$-integration} \label{beta}
For simplicity we write the double integral in Eq.\eqref{di} in three parts ,

\begin{itemize}
  \item \begin{eqnarray}
        T_1&=&\int\frac{d\alpha}{2\pi} \int\frac{d\beta}{2\pi}e^{-ik^+\alpha\beta(x_2^--x_1^-)} \left[{\rm{PV}}\left(\frac{1}{\beta}\right){\rm{PV}}\left(\frac{1}{1-\beta}\right)\frac{1}{(k_{1\perp}-\alpha k_\perp)^2}\right]\nn\\
        &&{l_\perp.(k_2-l)_\perp}\left[q_\perp.(l_\perp-\beta k_{2\perp})(k_{1\perp}-\alpha k_\perp).\epsilon_{\lambda_\perp}^*~-{\rm{PV}}\left(\frac{1}{(1-\alpha)}\right)q_\perp.(k_{1\perp}-\alpha k_\perp)~(l_\perp-\beta k_{2\perp}).\epsilon_{\lambda_\perp}^*\right.\nn\\
&&~~~~~~~~~~~~~~~~~~~~~~~~~~~~~~~\left.-{\rm{PV}}\left(\frac{1}{\alpha}\right)(k_{1\perp}-\alpha k_\perp).(l_\perp-\beta k_{2\perp})~q_\perp.\epsilon_{\lambda_\perp}^*\right]
  \end{eqnarray} 
  Performing $\beta-~$integration

  \begin{eqnarray}
         T_{11\beta}&=& \int\frac{d\beta}{2\pi}e^{-ik^+\alpha\beta(x_2^--x_1^-)} \left[{\rm{PV}}\left(\frac{1}{\beta}\right){\rm{PV}}\left(\frac{1}{1-\beta}\right)\right]
         q_\perp.(l_\perp-\beta k_{2\perp})\nn\\
         &=&\frac{i}{2}\left[q_\perp.l_\perp-q_\perp.(l_\perp-k_{2\perp})e^{-ik^+\alpha(x_2^--x_1^-)}\right]{\rm{Sign}}(x_2^--x_1^-)\nn
  \end{eqnarray}
  \begin{eqnarray}
         T_{12\beta}&=& \int\frac{d\beta}{2\pi}e^{-ik^+\alpha\beta(x_2^--x_1^-)} \left[{\rm{PV}}\left(\frac{1}{\beta}\right){\rm{PV}}\left(\frac{1}{1-\beta}\right)\right]
        (l_\perp-\beta k_{2\perp}).\epsilon_{\lambda_\perp}^*\nn\\
         &=&\frac{i}{2}\left[l_\perp.\epsilon_{\lambda_\perp}^*-(l_\perp-k_{2\perp}).\epsilon_{\lambda_\perp}^*e^{-ik^+\alpha(x_2^--x_1^-)}\right]{\rm{Sign}}(x_2^--x_1^-)\nn
  \end{eqnarray}
  \begin{eqnarray}
         T_{13\beta}&=& \int\frac{d\beta}{2\pi}e^{-ik^+\alpha\beta(x_2^--x_1^-)} \left[{\rm{PV}}\left(\frac{1}{\beta}\right){\rm{PV}}\left(\frac{1}{1-\beta}\right)\right]
        (k_{1\perp}-\alpha k_\perp).(l_\perp-\beta k_{2\perp})\nn\\
         &=&\frac{i}{2}\left[(k_{1\perp}-\alpha k_\perp).l_\perp-(k_{1\perp}-\alpha k_\perp).(l_\perp-k_{2\perp})e^{-ik^+\alpha(x_2^--x_1^-)}\right]{\rm{Sign}}(x_2^--x_1^-)\nn
  \end{eqnarray}

  with this $T_1$ becomes,\\

  \begin{empheq}{align*}
   T_1=&\int\frac{d\alpha}{2\pi}  \left[\frac{1}{(k_{1\perp}-\alpha k_\perp)^2}\right]\frac{i}{2}~{l_\perp.(k_2-l)_\perp}\nn\\
       & \left[(k_{1\perp}-\alpha k_\perp).\epsilon_{\lambda_\perp}^*\left\{q_\perp.l_\perp-q_\perp.(l_\perp-k_{2\perp})e^{-ik^+\alpha(x_2^--x_1^-)}\right\}~\right.\nn\\
        &\left.-{\rm{PV}}\left(\frac{1}{(1-\alpha)}\right)q_\perp.(k_{1\perp}-\alpha k_\perp)~\left\{l_\perp.\epsilon_{\lambda_\perp}^*-(l_\perp-k_{2\perp}).\epsilon_{\lambda_\perp}^*e^{-ik^+\alpha(x_2^--x_1^-)}\right\}\right.\nn\\
&~~~~~~\left.-{\rm{PV}}\left(\frac{1}{\alpha}\right)\left\{(k_{1\perp}-\alpha k_\perp).l_\perp-(k_{1\perp}-\alpha k_\perp).(l_\perp-k_{2\perp})e^{-ik^+\alpha(x_2^--x_1^-)}\right\}~q_\perp.\epsilon_{\lambda_\perp}^*\right]{\rm{Sign}}(x_2^--x_1^-)  
\end{empheq}
  
  Similarly for $T_2$ and $T_3$
   \item \begin{eqnarray}
        T_2&=&-\int\frac{d\alpha}{2\pi} \int\frac{d\beta}{2\pi}e^{-ik^+\alpha\beta(x_2^--x_1^-)} \left[{\rm{PV}}\left(\frac{1}{\beta}\right){\rm{PV}}\left(\frac{1}{1-\beta}\right)\frac{1}{(k_{1\perp}-\alpha k_\perp)^2}\right]\nn\\
&&{(l_\perp-\beta k_{2\perp}).(k_2-l)_\perp}{\rm{PV}}\left(\frac{1}{1-\beta}\right)\left[(k_{1\perp}-\alpha k_\perp).\epsilon_{\lambda_\perp}^*~l_\perp.q_\perp-{\rm{PV}}\left(\frac{1}{(1-\alpha)}\right)(k_{1\perp}-\alpha k_\perp).q_\perp~l_\perp.\epsilon_{\lambda_\perp}^*\right.\nn\\
&&~~~~~~~~~~~~~~~~~~~~~~~~~~~~~~~\left.-{\rm{PV}}\left(\frac{1}{\alpha}\right)(k_{1\perp}-\alpha k_\perp).l_\perp~q_\perp.\epsilon_{\lambda_\perp}^*\right]\nn
  \end{eqnarray} 
   Performing $\beta-~$integration
   \begin{eqnarray}
          T_{2\beta}&=& \int\frac{d\beta}{2\pi}e^{-ik^+\alpha\beta(x_2^--x_1^-)}\left[{\rm{PV}}\left(\frac{1}{\beta}\right){\rm{PV}}\left(\frac{1}{1-\beta}\right)\right]
          {(l_\perp-\beta k_{2\perp}).(k_2-l)_\perp}{\rm{PV}}\left(\frac{1}{1-\beta}\right)\nn\\
          &=&\frac{i}{2}\left[( k_{2\perp}-l_{\perp}).l_\perp-(k_{2\perp}-l_{\perp}).l_\perp e^{-ik^+\alpha(x_2^--x_1^-)}\right]{\rm{Sign}}(x_2^--x_1^-)\nn
   \end{eqnarray}
   
   with this $T_2$ becomes,\\

  \begin{empheq}{align*}
   T_2=&-\int\frac{d\alpha}{2\pi}  \left[\frac{1}{(k_{1\perp}-\alpha k_\perp)^2}\right]\frac{i}{2}~\left\{( k_{2\perp}-l_{\perp}).l_\perp-(k_{2\perp}-l_{\perp}).l_\perp e^{-ik^+\alpha(x_2^--x_1^-)}\right\}\nn\\
   &\left[(k_{1\perp}-\alpha k_\perp).\epsilon_{\lambda_\perp}^*~l_\perp.q_\perp-{\rm{PV}}\left(\frac{1}{(1-\alpha)}\right)(k_{1\perp}-\alpha k_\perp).q_\perp~l_\perp.\epsilon_{\lambda_\perp}^*\right.\nn\\
&~~~~~~~~~~~~~~~~~~~~~~~~~~~~~~~\left.-{\rm{PV}}\left(\frac{1}{\alpha}\right)(k_{1\perp}-\alpha k_\perp).l_\perp~q_\perp.\epsilon_{\lambda_\perp}^*\right]{\rm{Sign}}(x_2^--x_1^-)   
\end{empheq}

   \item \begin{eqnarray}
        T_3&=&-\int\frac{d\alpha}{2\pi} \int\frac{d\beta}{2\pi}e^{-ik^+\alpha\beta(x_2^--x_1^-)} \left[{\rm{PV}}\left(\frac{1}{\beta}\right){\rm{PV}}\left(\frac{1}{1-\beta}\right)\frac{1}{(k_{1\perp}-\alpha k_\perp)^2}\right]\nn\\
&&{(l_\perp-\beta k_{2\perp}).l_\perp}{\rm{PV}}\left(\frac{1}{\beta}\right)\left[(k_{1\perp}-\alpha k_\perp).\epsilon_{\lambda_\perp}^*~(k_2-l)_\perp.q_\perp-{\rm{PV}}\left(\frac{1}{(1-\alpha)}\right)(k_{1\perp}-\alpha k_\perp).q_\perp~(k_2-l)_\perp.\epsilon_{\lambda_\perp}^*\right.\nn\\
&&~~~~~~~~~~~~~~~~~~~~~~~~~~~~~~~\left.-{\rm{PV}}\left(\frac{1}{\alpha}\right)(k_{1\perp}-\alpha k_\perp).(k_2-l)_\perp~q_\perp.\epsilon_{\lambda_\perp}^*\right] \nn    
  \end{eqnarray} 
  Performing $\beta-~$integration
   \begin{eqnarray}
          T_{3\beta}&=& \int\frac{d\beta}{2\pi}e^{-ik^+\alpha\beta(x_2^--x_1^-)}\left[{\rm{PV}}\left(\frac{1}{\beta}\right){\rm{PV}}\left(\frac{1}{1-\beta}\right)\right]
          {(l_\perp-\beta k_{2\perp}).l_\perp}{\rm{PV}}\left(\frac{1}{\beta}\right)\nn\\
          &=&\frac{i}{2}\left[(l_{\perp}- k_{2\perp}).l_\perp-(l_{\perp}- k_{2\perp}).l_\perp e^{-ik^+\alpha(x_2^--x_1^-)}\right]{\rm{Sign}}(x_2^--x_1^-)\nn
   \end{eqnarray}
  
     with this $T_3$ becomes,\\

  \begin{empheq}{align*}
   T_3=&-\int\frac{d\alpha}{2\pi}  \left[\frac{1}{(k_{1\perp}-\alpha k_\perp)^2}\right]\frac{i}{2}~\left\{(l_{\perp}- k_{2\perp}).l_\perp-(l_{\perp}- k_{2\perp}).l_\perp e^{-ik^+\alpha(x_2^--x_1^-)}\right\}\nn\\
   &\left[(k_{1\perp}-\alpha k_\perp).\epsilon_{\lambda_\perp}^*~(k_2-l)_\perp.q_\perp-{\rm{PV}}\left(\frac{1}{(1-\alpha)}\right)(k_{1\perp}-\alpha k_\perp).q_\perp~(k_2-l)_\perp.\epsilon_{\lambda_\perp}^*\right.\nn\\
&~~~~~~~~~~~~~~~~~~~~~~~~~~~~~~~\left.-{\rm{PV}}\left(\frac{1}{\alpha}\right)(k_{1\perp}-\alpha k_\perp).(k_2-l)_\perp~q_\perp.\epsilon_{\lambda_\perp}^*\right]{\rm{Sign}}(x_2^--x_1^-)   
\end{empheq}
   
\end{itemize}

  \section{$\alpha$-integration} \label{alpha}      
         
     Writing expression for amplitude in two parts, one without exponential ($T_{a}$) and other with exponential ($T_{b}$).\\
\begin{eqnarray}
  T_{a}&=&\int\frac{d\alpha}{2\pi}  \left[\frac{-\left((l_{\perp}- k_{2\perp}).l_\perp\right)}{(k_{1\perp}-\alpha k_\perp)^2}\right]\frac{i}{2}~{\rm{Sign}}(x_2^--x_1^-)\nn\\  
  &&\left[(k_{1\perp}-\alpha k_\perp).\epsilon_{\lambda_\perp}^*\left\{(k_2-l)_\perp.q_\perp\right\}-{\rm{PV}}\left(\frac{1}{1-\alpha}\right)\left\{(k_{1\perp}-\alpha k_\perp).q_\perp~(k_2-l)_\perp.\epsilon_{\lambda_\perp}^*
 \right\}\right.\nn\\
 &&\left.-{\rm{PV}}\left(\frac{1}{\alpha}\right)\left\{(k_{1\perp}-\alpha k_\perp).(k_2-l)_\perp~q_\perp.\epsilon_{\lambda_\perp}^*
 \right\}
  \right]\nn
\end{eqnarray}
Solving $T_a$,
\begin{itemize}
    \item \begin{eqnarray}
           T_{a\alpha_1}&=&{l_\perp.(k_2-l)_\perp}~(k_2-l)_\perp.q_\perp\frac{i}{2}\int\frac{d\alpha}{2\pi}  \left[\frac{1}{(k_{1\perp}-\alpha k_\perp)^2}\right]~{\rm{Sign}}(x_2^--x_1^-)\left[(k_{1\perp}-\alpha k_\perp).\epsilon_{\lambda_\perp}^*\right]\nn\\
           &=&{l_\perp.(k_2-l)_\perp}~(k_2-l)_\perp.q_\perp\frac{i}{2} {\rm{Sign}}(x_2^--x_1^-)\left[\frac{\epsilon_{\lambda_\perp}^*\left(k_\perp^2k_{1\perp}-(k_{1\perp}.k_\perp)k_\perp\right)}{k_{\perp}^2|k_\perp\times k_{1\perp}|}\right]\nn\\
           &=&{l_\perp.(k_2-l)_\perp}~(k_2-l)_\perp.q_\perp\frac{i}{2} {\rm{Sign}}(x_2^--x_1^-)\left[\frac{-\epsilon_{\lambda_\perp}^*\times k_\perp~k_\perp\times k_{1\perp}}{k_{\perp}^2|k_\perp\times k_{1\perp}|}\right]
\end{eqnarray}

\item \begin{eqnarray}
           T_{a\alpha_2}&=&-{l_\perp.(k_2-l)_\perp}\frac{i}{2}\int\frac{d\alpha}{2\pi}  \left[\frac{1}{(k_{1\perp}-\alpha k_\perp)^2}\right]~{\rm{Sign}}(x_2^--x_1^-){\rm{PV}}\left(\frac{1}{(1-\alpha)}\right)\left\{q_\perp.(k_{1\perp}-\alpha k_\perp)~(k_2-l)_{\perp}.\epsilon_{\lambda_\perp}^*
 \right\}\nn\\
           &=&-{l_\perp.(k_2-l)_\perp}(k_2-l)_{\perp}.\epsilon_{\lambda_\perp}^*\frac{i}{2} {\rm{Sign}}(x_2^--x_1^-)\left[\frac{q_\perp.\left\{k_\perp~k_{1\perp}.(k_\perp-k_{1\perp})-k_{1\perp}k_\perp.(k_\perp-k_{1\perp})\right\}}{(k_\perp-k_{1\perp})^2|k_\perp\times k_{1\perp}|}\right]\nn\\
           &=&-{l_\perp.(k_2-l)_\perp}(k_2-l)_{\perp}.\epsilon_{\lambda_\perp}^*\frac{i}{2} {\rm{Sign}}(x_2^--x_1^-)\left[\frac{q_\perp\times(k_\perp-k_{1\perp})~k_\perp\times k_{1\perp}}{(k_\perp-k_{1\perp})^2|k_\perp\times k_{1\perp}|}\right]\nn
\end{eqnarray}
\item \begin{eqnarray}
           T_{a\alpha_3}&=&-{l_\perp.(k_2-l)_\perp}~q_\perp.\epsilon_{\lambda_\perp}^*\frac{i}{2}\int\frac{d\alpha}{2\pi}  \left[\frac{1}{(k_{1\perp}-\alpha k_\perp)^2}\right]~{\rm{Sign}}(x_2^--x_1^-)\left[{\rm{PV}}\left(\frac{1}{\alpha}\right)\left\{(k_2-l)_\perp.(k_{1\perp}-\alpha k_\perp) \right\}\right]\nn\\
           &=&-{l_\perp.(k_2-l)_\perp}~q_\perp.\epsilon_{\lambda_\perp}^*\frac{i}{2} {\rm{Sign}}(x_2^--x_1^-)\left[\frac{(k_2-l)_\perp.\left\{k_\perp~k_{1\perp}^2-k_{1\perp}(k_\perp.k_{1\perp})\right\}}{k_{1\perp}^2|k_\perp\times k_{1\perp}|} \right]\nn\\
           &=&-{l_\perp.(k_2-l)_\perp}~q_\perp.\epsilon_{\lambda_\perp}^*\frac{i}{2} {\rm{Sign}}(x_2^--x_1^-)\left[\frac{(k_2-l)_\perp\times k_{1\perp}~k_\perp\times k_{1\perp}}{k_{1\perp}^2|k_\perp\times k_{1\perp}|} \right]\nn
\end{eqnarray}

\end{itemize}

\begin{empheq}[box=\fbox]{align*}
   T_a=&{l_\perp.(k_2-l)_\perp}~\frac{i}{2} {\rm{Sign}}(x_2^--x_1^-){\rm{Sign}}(k_\perp\times k_{1\perp})\left[(k_2-l)_\perp.q_\perp\frac{-\epsilon_{\lambda_\perp}^*\times k_\perp}{k_{\perp}^2}\right.\nn\\
   &\left.-(k_2-l)_{\perp}.\epsilon_{\lambda_\perp}^*\frac{q_\perp\times(k_\perp-k_{1\perp})}{(k_\perp-k_{1\perp})^2}-q_\perp.\epsilon_{\lambda_\perp}^*\frac{(k_2-l)_\perp\times k_{1\perp}}{k_{1\perp}^2} \right]
\end{empheq}

\begin{eqnarray}
  T_{b}&=&\int\frac{d\alpha}{2\pi}  \left[\frac{{l_\perp.(k_2-l)_\perp}}{(k_{1\perp}-\alpha k_\perp)^2}\right]\frac{i}{2}~e^{-ik^+\alpha(x_2^--x_1^-)}{\rm{Sign}}(x_2^--x_1^-)\nn\\  
  &&\left[(k_{1\perp}-\alpha k_\perp).\epsilon_{\lambda_\perp}^*~l_\perp.q_\perp-{\rm{PV}}\left(\frac{1}{(1-\alpha)}\right)\left\{q_\perp.(k_{1\perp}-\alpha k_\perp)~l_{\perp}.\epsilon_{\lambda_\perp}^*
 \right\}-{\rm{PV}}\left(\frac{1}{\alpha}\right)\left\{l_\perp.(k_{1\perp}-\alpha k_\perp)~q_\perp.\epsilon_{\lambda_\perp}^*
 \right\}
  \right]\nn
\end{eqnarray}
Solving $T_b$,
\begin{itemize}
    \item \begin{eqnarray}
           T_{b\alpha_1}&=&{l_\perp.(k_2-l)_\perp}~l_\perp.q_\perp\frac{i}{2}\int\frac{d\alpha}{2\pi}  \left[\frac{1}{(k_{1\perp}-\alpha k_\perp)^2}\right]~e^{-ik^+\alpha(x_2^--x_1^-)}{\rm{Sign}}(x_2^--x_1^-)\left[(k_{1\perp}-\alpha k_\perp).\epsilon_{\lambda_\perp}^*\right]\nn\\
           &=&{l_\perp.(k_2-l)_\perp}~l_\perp.q_\perp\frac{i}{2} {\rm{Sign}}(x_2^--x_1^-)\left[\frac{\epsilon_{\lambda_\perp}^*\left(k_\perp^2k_{1\perp}-(k_{1\perp}.k_\perp)k_\perp\right)}{k_{\perp}^2|k_\perp\times k_{1\perp}|}+i\frac{{\rm{Sign}}(x_2^--x_1^-)|k_\perp\times k_{1\perp}|}{k_\perp^2|k_\perp\times k_{1\perp}|} \right]\nn\\
           &=&{l_\perp.(k_2-l)_\perp}~l_\perp.q_\perp\frac{i}{2} {\rm{Sign}}(x_2^--x_1^-)\left[\frac{-\epsilon_{\lambda_\perp}^*\times k_\perp~k_\perp\times k_{1\perp}}{k_{\perp}^2|k_\perp\times k_{1\perp}|}+i\frac{{\rm{Sign}}(x_2^--x_1^-)}{k_\perp^2} \right]
\end{eqnarray}
\item \begin{eqnarray}
           T_{b\alpha_2}&=&-{l_\perp.(k_2-l)_\perp}\frac{i}{2}\int\frac{d\alpha}{2\pi}  \left[\frac{1}{(k_{1\perp}-\alpha k_\perp)^2}\right]~e^{-ik^+\alpha(x_2^--x_1^-)}{\rm{Sign}}(x_2^--x_1^-){\rm{PV}}\left(\frac{1}{(1-\alpha)}\right)\left\{q_\perp.(k_{1\perp}-\alpha k_\perp)~l_{\perp}.\epsilon_{\lambda_\perp}^*
 \right\}\nn\\
           &=&-{l_\perp.(k_2-l)_\perp}l_{\perp}.\epsilon_{\lambda_\perp}^*\frac{i}{2} {\rm{Sign}}(x_2^--x_1^-)\left[\frac{q_\perp.\left\{k_\perp~k_{1\perp}.(k_\perp-k_{1\perp})-k_{1\perp}k_\perp.(k_\perp-k_{1\perp})\right\}}{(k_\perp-k_{1\perp})^2|k_\perp\times k_{1\perp}|}\right]\nn\\
            &=&-{l_\perp.(k_2-l)_\perp}l_{\perp}.\epsilon_{\lambda_\perp}^*\frac{i}{2} {\rm{Sign}}(x_2^--x_1^-)\left[\frac{q_\perp\times (k_\perp-k_{1\perp})~k_\perp\times k_{1\perp}}{(k_\perp-k_{1\perp})^2|k_\perp\times k_{1\perp}|}\right]\nn
\end{eqnarray}
\item \begin{eqnarray}
           T_{b\alpha_3}&=&-{l_\perp.(k_2-l)_\perp}~q_\perp.\epsilon_{\lambda_\perp}^*\frac{i}{2}\int\frac{d\alpha}{2\pi}  \left[\frac{1}{(k_{1\perp}-\alpha k_\perp)^2}\right]~e^{-ik^+\alpha(x_2^--x_1^-)}{\rm{Sign}}(x_2^--x_1^-)\left[{\rm{PV}}\left(\frac{1}{\alpha}\right)\left\{l_\perp.(k_{1\perp}-\alpha k_\perp) \right\}\right]\nn\\
           &=&-{l_\perp.(k_2-l)_\perp}~q_\perp.\epsilon_{\lambda_\perp}^*\frac{i}{2} {\rm{Sign}}(x_2^--x_1^-)\left[\frac{l_\perp.\left\{k_\perp~k_{1\perp}^2-k_{1\perp}(k_\perp.k_{1\perp})\right\}}{k_{1\perp}^2|k_\perp\times k_{1\perp}|} \right]\nn\\
           &=&-{l_\perp.(k_2-l)_\perp}~q_\perp.\epsilon_{\lambda_\perp}^*\frac{i}{2} {\rm{Sign}}(x_2^--x_1^-)\left[\frac{l_\perp\times k_{1\perp}~k_{\perp}\times k_{1\perp}}{k_{1\perp}^2|k_\perp\times k_{1\perp}|} \right]\nn
\end{eqnarray}

\begin{empheq}[box=\fbox]{align*}
   T_b=&{l_\perp.(k_2-l)_\perp}~\frac{i}{2} {\rm{Sign}}(x_2^--x_1^-)\left[i~l_\perp.q_\perp~\frac{{\rm{Sign}}(x_2^--x_1^-)}{k_\perp^2}+{\rm{Sign}}(k_\perp\times k_{1\perp})\left\{l_\perp.q_\perp\frac{-\epsilon_{\lambda_\perp}^*\times k_\perp}{k_{\perp}^2}\right.\right.\nn\\
   &\left.\left.-l_{\perp}.\epsilon_{\lambda_\perp}^*\frac{q_\perp\times(k_\perp-k_{1\perp})}{(k_\perp-k_{1\perp})^2}-q_\perp.\epsilon_{\lambda_\perp}^*\frac{(k_2-l)_\perp\times k_{1\perp}}{k_{1\perp}^2} \right\}\right]\nn
\end{empheq}
\end{itemize}

\begin{eqnarray}
       T_a+T_b&=&{-l_\perp.(k_2-l)_\perp}~\frac{i}{2} {\rm{Sign}}(x_2^--x_1^-)\left[k_{2\perp}.q_\perp\frac{\epsilon_{\lambda_\perp}^*\times k_\perp}{k_{\perp}^2}\right.\nn\\
   &&\left.+k_{2\perp}.\epsilon_{\lambda_\perp}^*\frac{q_\perp\times(k_\perp-k_{1\perp})}{(k_\perp-k_{1\perp})^2}+2q_\perp.\epsilon_{\lambda_\perp}^*\frac{(k_2-l)_\perp\times k_{1\perp}}{k_{1\perp}^2} \right]{\rm{Sign}}(k_\perp\times k_{1\perp})-{l_\perp.(k_2-l)_\perp}~\frac{l_\perp.q_\perp}{2k_\perp^2}
\end{eqnarray}


 \end{document}